\documentclass[a4paper,12pt]{article}

\usepackage{amssymb,amsmath,mathbbol,mathrsfs}
\usepackage[usenames,dvipsnames]{color}
\usepackage{hyperref}
\usepackage{xcolor}
\usepackage{stmaryrd}
\usepackage{authblk}
\usepackage{framed}
\usepackage{empheq}
\usepackage{slashed}
\usepackage{hyphenat}
\usepackage{apacite}
\usepackage{enumerate}

\usepackage{parskip}

\usepackage{marginnote}    

\usepackage[left=.55in,right=.55in,top=.8in,bottom=.7in]{geometry}                  

\usepackage{sectsty}
\allsectionsfont{\sffamily\mdseries\upshape} 
\usepackage{tocloft}

\makeatletter
\renewenvironment{abstract}{%
    \if@twocolumn
      \section*{\abstractname}%
    \else 
      \begin{center}%
        {\bfseries\sffamily\abstractname\vspace{\z@}}
      \end{center}%
      \quotation
    \fi}
    {\if@twocolumn\else\endquotation\fi}
\makeatother

5

\hypersetup{
	colorlinks=true,         
	linkcolor=MidnightBlue,          
	citecolor=BrickRed,        
	urlcolor=MidnightBlue            
}

\newcommand{\be}{\begin{equation}}
\newcommand{\ee}{\end{equation}}

\newcommand{\F}{{\Phi}}
\newcommand{\RR}{\mathbb{R}} 
\renewcommand{\d}{{\mathrm{d}}}
\newcommand{\D}{{\mathrm{D}}}

\newcommand{\G}{{\mathcal{G}}}

\renewcommand{\bar}{\overline}

\newcommand{\cint}{{\int\kern-.87em{<}}}
\newcommand{\sint}{{\int\kern-.75em{\sim}}}
\newcommand{\fint}{{\int\kern-1.00em{\int}}}

\let\oldmarginpar\marginpar
\renewcommand\marginpar[1]{\oldmarginpar{\color{red}\raggedright\footnotesize #1}}

\begin{document}
\title{Boundaries, frames and the issue of physical covariance}
\author[1]{ Henrique Gomes\thanks{henrique.deandradegomes@philosophy.ox.ac.uk}}
\author[2]{Simon Langenscheidt \thanks{S.Langenscheidt@physik.lmu.de}}
\author[3]{ Daniele Oriti \thanks{doriti@ucm.es}}
\affil[1]{\small Oriel College, Oxford, UK OX1 4EW}
\affil[2]{\small Arnold Sommerfeld Center for Theoretical Physics and MCQST, Ludwig-Maximilians University, Munich, Germany, EU}
\affil[3]{\small Depto. de Física Teórica and IPARCOS, Facultad de Ciencias Físicas \\
Universidad Complutense de Madrid, Spain, EU}

\maketitle

\abstract{We focus on three distinct lines of recent developments: edge modes and boundary charges in gravitational physics, relational dynamics in classical and quantum gravity, and quantum reference frames. Recollecting a now well-established set of results, we emphasize that these research directions are in fact linked in multiple ways, and can be seen as different aspects of the same research programme. This research programme has two main physical goals and one general focus, as well as broader conceptual implications. The physical goals are to move beyond the two idealizations/approximations of asymptotic or closed boundary conditions in gravitational physics and of ideal reference frames (coded in coordinate frames or gauge fixings), thus achieving a more realistic modelling of (quantum) gravitational physical phenomena. These two goals combine to identify a key open issue: a proper characterization of physical covariance, i.e. covariance across fully physical (as opposed to idealized) reference frames. The broader conceptual implications concern the influence of observers in physics and possible physical limits to objectivity. 
 }


\section*{Introduction}
\setlength{\parindent}{0pt}
\setlength{\parskip}{0.2em}

In this contribution, we focus on two kinds of idealizations that are routinely employed when modeling physics systems in a gravitational context. Namely, that of negligible spacetime boundaries (often associated with the restriction to closed systems) and that of ideal reference frames. In this paper, we will investigate what happens upon the removal of these idealizations, i.e. when we make our models more realistic.

\noindent In doing so, we think more carefully about the modelling process itself, and specifically take seriously recent developments in classical and quantum gravitational physics as well as quantum foundations, pointing towards some of their non-trivial implications. These have to do with gravitational (more generally, gauge) systems in finite spacetime regions, relational dynamics in classical and quantum gravity, and quantum reference frames. Lastly, we will tie together the main lessons we gather from such developments and conclude that, when the above idealizations are removed, we are left with a description of gravitational systems which is necessarily perspectival through a dependence on the adopted physical reference frame. In turn, this is understood as the (partial) embodiment of the agent modelling the system or (depending on the context) the observer accessing it to determine its spatiotemporal properties.\footnote{The  notions of agent and observer are obviously distinct, and modelling a system is not the same as observing it; however, in physics we model systems with the idea that their properties can be at least indirectly \lq observed\rq or empirically accessed by their physical consequences, and in spacetime physics, which is our present focus, this \lq access\rq to the system includes a determination of its location in space and time. Thus an epistemic agent modelling a system would also be \lq observing\rq the system when specifying its location in space and time. Thus in the following we often use either notion, depending on the aspects of the combined modelling/observing activity we want to emphasize.} We also emphasize that this `perspectivalism' has the consequence that relating the descriptions made by different observers is no longer a straightforward matter. 


\noindent Before we delve into our main arguments and the conclusion they lead us to, we introduce a bit of context and broadly motivate our reasoning and our attention to the issue of boundaries and frames in physics, and gravitational theory in particular.

We are sympathetic to operational attitudes toward physical modeling and believe that physical models should, as much as possible, reflect the realistic conditions of our interactions with the relevant physical systems. We consider this an important component, albeit not the only relevant one, of scientific methodology. 


We understand the modelling of the system as an activity performed by a variety of types of agents \cite{barzegar_minimalist_2023}. For this activity to make sense, there must be an initial separation between the agent, including any equipment relevant to the modelling activity (recording devices, etc.), and the system that is being modelled by the agent. Such a split must also be reasonably stable under the dynamics of both agent and modelled system.

Particularly, among their equipment, we focus on anything used to define a reference frame for spatiotemporal properties: other physical systems that the agent uses to project properties of the modelled system into a form that could encode its spacetime localization, e.g. a set of chosen clock and rods. 

These reference frames serve, of course, also the purpose of contextualising the agent itself within the larger system it constitutes together with the modelled system. It is through their reference frame that an agent acquires a perspective on the full system and its properties; one issue we will discuss is whether the reference frame itself affects said perspective in ways that are ultimately non-negligible.

A central focus of our discussion is how  physical systems defining a reference frames are actually modelled for a specific gravitational system (which could be the spacetime geometry itself), and how many of their own physical properties are accounted for in the model.

Indeed, one can (and, we will argue, should) include such a reference frame explicitly within the model of the system of interest. The issue becomes then how this inclusion, when performed in less idealized terms, changes the description of the modelled system and what the conceptual implications of these changes are.

A second, related way in which the adoption of an agent/observer-based perspective influences our modelling of gravitational systems has to do with the resulting localization of the modelled system in spacetime, regardless (initially) of how this localization is determined via reference frames. 

The modelled system can be seen as occupying a specific portion of spacetime, and this portion does not include the observer/agent herself, i.e. it is not the same portion of spacetime that the observer/agent occupies (as she does not appear in the model of the system). The observer/system split, therefore, is also reflected in the existence of a boundary of spacetime, separating the two.
These boundaries typically take the form of some timelike or lightlike submanifolds which enclose either the observer or the system (up to being possibly 'capped off' by some spacelike initial or final surfaces).\\
Neglecting the role of agents/observers and the existence of such split translates, in our models, into the treatment of the system as closed and also in its description as occupying a dynamically closed region of spacetime, modelled as a closed manifold without boundary, or as a manifold with boundary but with special boundary conditions preventing information or degrees of freedom from leaving or entering the system. The inclusion of spacetime boundaries in our models, moreover, can be done in more realistic or more idealized terms, and the ensuing question is, again, how removing idealizations affects the act of modelling itself. 

So what happens when, in our models, we do not neglect the presence of spacetime boundaries? As we will argue, their presence is relevant as they do affect the physics of the systems of interest; this is also closely tied to the issue of modelling, in a less idealized way, physical reference frames. 

In the case of spacetime boundaries, the idealizations involved are of two types. The more drastic one is to simply neglect the existence of boundaries and treat the system of interest as closed even in its spacetime and gravitational properties, as we discussed. Less drastic, but still an idealization, is the assumption that the spacetime boundary (thus, the observer) can be \lq pushed to infinity\rq, i.e. the imposition of asymptotic boundary conditions and of infinite distance/volume limits. In a gravitational context, this is often accompanied by the switching off of the gravitational interaction between the system of interest and the system \lq outside the boundary\rq, when one imposes flat or weak gravity data at the asymptotic boundary. The growing interest in gravitational systems in finite regions has resulted in a large number of recent developments, which in turn have improved also our understanding of the asymptotic setting, with both foundational and possibly observational consequences. Moreover, the physics of gravitational systems in finite regions ties closely to the physics of reference frames and relational dynamics, as we will discuss in the following.

We therefore arrive at our main point. A more realistic understanding of coupling of (gravitational) systems, going beyond typical idealizations, and the construction of physical questions, as represented by gauge invariant observables, points to a non-negligible role of physical frames (as partial embodiment of observers) in the understanding of gravitational systems, at classical and even more so at quantum levels. \\
Furthermore, this has implications on what we should expect in terms of comparison and translation of statements made by different observers, and it suggests an irreducible perspectival nature of such a physical understanding.

We will now present in more detail several arguments for these ideas, and give pointers to relevant ongoing research directions before discussing briefly some of the broader conclusions we draw from them.

\section*{The issue of physical (non)covariance}
Drawing on a solid body of results from mathematical gravitational physics, quantum gravity and foundations of physics, the main conclusion we want to argue for, in this contribution, is the following.

Once we remove two standard idealizations from our models of gravitational systems, whether classical or quantum, and in particular, once we take seriously the distinction between open and closed systems (encoded in the most realistic way in the presence of finite spacetime boundaries), as well as the need for physical reference frames (encoded in gauge invariance and nonlocality with respect to the supporting manifold), we realize that a crucial issue remains open in fundamental gravitational physics: that of physical covariance with respect to transformations across physical, realistic reference frames, together with the question of what, exactly, is in fact invariant under the same transformations.
Our reasoning points to the inevitability of this issue, and thus calls for devoting to it our full research attention.



\


Before presenting the main steps in our argument, and their supporting evidence,  we state the basic presumptions we are starting from. 

\begin{enumerate}[I.]

    \item Modelling conventionally starts with a system/agent or system/observer split. Modelling a physical system requires one to first identify all the data and observables referring to the system and then setting up an epistemic boundary between the system itself and the agent performing the modelling. By such a boundary, we mean both a conceptual or abstract split of all of the degrees of freedom into independent parts that are, to a certain extent, controllable by the agent, as well as any concrete physical object or phenomenon that implements such a split on the level of the actual system. This provides some separation of the agent from the system to be modelled,  i.e. for all properties whose exclusion does not impair the explanatory or predictive accuracy of the model.\\ Such properties may, however, as we will argue, possibly include some properties of the agents themselves. 
    Therefore, we presume that the split is at first more a matter of convenience, and that the agent can, within reason, include their own properties and dynamics in the model of the system, when these are considered relevant. 
    If the system is to be also modelled in its spatiotemporal properties---and this is clearly necessarily the case for gravitational systems and, in fact, extends to any dynamical system---its spacetime localization is encoded by assigning it a determined portion of the manifold we conventionally use to represent spacetime. The existence of a system/observer split is then encoded in the presence of a boundary for such a region, with this boundary being finite in realistic models (and before any further approximation). This is the case both if the model represents the observer as occupying a finite spacetime region, with the system constituting whatever is 'left over'; or, more customary, if it is the system that is being modelled as occupying a finite spacetime region (the observer being ascribed as part of \lq the environment\rq).
    \item We also assume that modelling activity may involve a general type of reference frame. By a reference frame, we generally refer to any, possibly entirely conceptual, entities used as a way by which an agent standardises their recordings of phenomena in the system. The simplest example of this, in the context of spacetime physics and gravitational systems, is a coordinate frame, which consists of a set of conventional real labels the agent assigns (in principle arbitrarily) to the points in the manifold mentioned before, interpreted as an encoding of the temporal and spatial localization of the \lq event associated with that point\rq. For instance,  a model of the reading of some clock and rods. By fixing such a frame, the agent can organise their recordings by projecting them into a set of numbers respective to the frame, for example the mentioned coordinates. We can speak further of a reference frame as \textit{physical}, which we understand as any set of agent-controllable or accessible degrees of freedom which can serve as a means to situate the agent within the full system, and serve as a standard to record measurements about the system (thus, they are frames according to the general definition), but are modelled by accounting explicitly for the dynamical and physical properties of the actual subsystems used by the agent to measure/access (at least indirectly or hypothetically) the system. In the case of clock and rods, of interest for what follows, they translate in the model into a physical reference frame to the extent in which their actual physical properties as matter systems are accounted for. 
    \item It is possible, interesting, and often necessary, to include aspects of the observers/agents in the model of the system of interest. One could then say that the abstract observer/agent has been \lq embodied\rq\, in specific physical systems or mathematical conditions (e.g. boundary conditions, external potentials, etc) appearing in the model of the system of interest. 
    In particular, an aspect of observers/agents that can be embodied in the model  corresponds exactly to the spatiotemporal reference frames, i.e. the set of clock and rods used to localize the physical system of interest in space and time (in fact, one could say that these devices \textit{define} such localization). This is where the issue of how realistically reference frames are modelled becomes relevant, and with it, in particular, the distinction between coordinate and physical reference frames. This distinction will be a key point in our argument below.
    To anticipate aspects of our argument, if reference frames are to be considered part of physical reality, and not just abstract elements in the modelling process, their physical properties, including their interaction with other physical systems and their backreaction on spacetime geometry, should be carefully modelled and quantified (in particular if they are to be modelled within a theory of quantum gravity and quantum spacetime). At the quantum level, the quantum properties of the physical systems used as reference frames should also be accounted for in the model. 
    \end{enumerate}

 \
 
Now, on the basis of these presumptions, let us present our arguments in a way that builds up to the conclusion we want to emphasize. 

\

\noindent \textbf{0.}
\textbf{Starting point}\\
The localization of physical systems in spacetime and their dynamical evolution through it are conventionally modelled in terms of points and trajectories in a differentiable manifold, and made explicit, in each region of such a manifold, by their expression in terms of coordinate frames. Their domain of definition identifies the spacetime regions which can constitute the domain of the physical systems under consideration, often assumed to run up to infinity in time and/or space. 
 Both the association of coordinate frames with observers (as they are defined/assigned by them), or, better, the modelling of the reference frames used by the observers as coordinate frames, and their extending to infinity encode two important idealizations, on whose removal from our models we focus in the following.

  \
Steps 1 and 2 will contain most of the technical details required to make the remaining arguments, and are for that reason longer and more detailed. 

\

\textbf{1.}
\textbf{Diffeomorphism invariance implies that coordinate frames are unphysical}

Conceptually, the gauge symmetries of gravitational theories, specifically diffeomorphism invariance, leads to the view that coordinate frames are entirely unphysical: they are elements of the formalism (of manifolds and modelling) that do not affect physical results and that have no physical significance.  Of course, the use of coordinates has enormous practical advantages but should be recognized as not developing a foundational role in our understanding of gravitational systems (and by extension, of the physical world). In fact, full invariance under active diffeomorphisms runs deeper than just coordinate invariance \cite{GaulRovelli}, and it has important consequences for the dynamical aspects of spacetime in GR-like theories \cite{Giulini}.

Whether this lack of physical status of coordinate reference frames extends to points and trajectories in the manifold, and thus to the manifold itself, is more subtle. One could hold the view that what is physical must be gauge-invariant. 
In this light, one concludes that bare manifold points (thus directions, subregions, etc) carry no physical significance, since there are diffeomorphisms which are symmetries of the theory and won't preserve the given subregions. However, even adopting this view, one may still wonder about the status of quantities that enter into the definition of gauge-invariant observables; for instance, field values whose relations to other field values may together constitute a gauge-invariant observable \cite{RovelliPartialComplete}. 
Since no function  that carries manifold points as free (or unbound) variables 
is, in itself, a gauge-invariant observable (diffeomorphisms map points to other points and thus transform non-trivially such functions), and this includes any set of field values, such parts of an invariant observable need not be themselves invariant: they only need to have the right covariance property.  Thus all invariant observables need to be instantiated by suitable diffeomorphism-invariant functions of field values \cite{Dittrich,goeller_diffeomorphism-invariant_2022}; functions in which the bare manifold points either don't appear or figure only as bound variables (in this limited sense, reference to manifold points can be considered unproblematic). Such functions could be the relational observables we discuss below.

In any case, it is conceivable that manifold points, directions, paths etc may altogether disappear from the formulation of our models, and may not feature at all in more fundamental theories, i.e. in quantum gravity. Explicit examples of such a disappearance are given in classical GR, when this is deparametrized in terms of special material frames \cite{BrownKuchar,GieselThiemann}, 
and in a handful of approaches to quantum gravity, in various approximations \cite{MarchettiOriti,GieselThiemannScalar}, and in toy models like those that appear in quantum cosmology \cite{LQC}.

Technically, we can associate these 
issues regarding observables to the mathematical formulation of gauge theories as follows. 
In gauge theories and in gravity, dynamical symmetries of the basic variables involve spatiotemporal derivatives of arbitrary functions. It follows that the symmetry-invariant content of the initial state can be extracted only non-locally from the basic variables, from procedures that involve spacetime points only as bound variables, e.g. as variables being integrated over. In more detail, a local symmetry implies, via Noether\rq{}s second theorem (cf. \cite{BradingBrown_Noether} for a conceptual exposition), that the equations of motion of the theory are not all independent: the full set of Euler-Lagrange equations is not linearly independent, or, alternatively, it obeys constraints. Thus this set of equations only \textit{uniquely} determines the evolution of a subset of the original degrees of freedom. This subset can be interpreted as made of `composite' variables: it is described by relations between the original degrees of freedom. The evolution of the remaining degrees of freedom is arbitrary. These remaining degrees of freedom are usually taken to not describe any physical feature of the system and are called \textit{gauge}. There is thus a certain freedom in choosing which composites of, or relations between, the original fields will be uniquely propagated, or will evolve deterministically---each choice corresponds to  \emph{a choice of gauge}. 

In the Hamiltonian phase space formalism, this indeterminism comes in the form of first-class constraints that physical observables have to satisfy. A choice of physical degrees of freedom is a choice of canonical degrees of freedom that satisfy the constraints, which amount to choices of phase space functions that completely fix the gauge freedom (and, in Hamiltonian analysis, are phase space functions that form a 'second-class' system together with the constraints). In the next item, we will see how these technical features are associated to a (benign) form of non-locality.

\

\textbf{2.}
\textbf{Localization in space  and relational time requires physical frames.}

The implication of gauge symmetries for observables aligns, in fact, with a more realistic (and operational) description of spacetime localization and evolution. Observers use physical (material) clocks and rods to define them, so physical observables are best understood as relations between dynamical fields and the (material) clock and rods used for their localization in spacetime. This is the relational strategy for the definition of physical, thus diffeomorphism-invariant, observables in gravitational physics \cite{RovelliPartialComplete,Dittrich}. Their construction is often difficult, but their in principle availability is a fact. It is a fact, that is, provided one seeks them to be defined on subregions of the supporting differentiable manifold only, and not globally, and accepts the need to use several physical frames for a global description of physical systems throughout space and time. State-of-the-art explicit construction for this can be found in \cite{goeller_diffeomorphism-invariant_2022}. \\
Explicit examples of such relational constructions are given in classical GR, deparametrised in terms of special physical frames \cite{BrownKuchar,GieselThiemann}, 
in a handful of approaches to quantum gravity, in various approximations \cite{MarchettiOriti,GieselThiemannScalar}, especially in toy models like in quantum cosmology \cite{LQC,hoehn_how_2020}.

One way to find such descriptions in practice, what we alluded to above as `choosing a gauge', is also known as  \textit{gauge fixing}. 
Gauge-fixing functions can often be interpreted as characterising a \textit{reference frame} for the gauge group in question. E.g. in Newtonian mechanics, a gauge-fixing of translations is given by fixing the center of mass to be immobile, and this fixes the frame to be the center of mass frame, which is a thoroughly relational entity.

Note that the Newtonian example does not illustrate the main feature that we are describing here: indeterminism of the equations of motion. The reason is that classical non-relativistic particle mechanics does not admit time-dependent symmetries and so does not suffer from indeterminism (see, e.g., \cite{Wallace_LagSym}. Nonetheless, the example is similar in the relevant aspects to gauge-fixing in general relativity or Yang-Mills theory.

Technically, the composite or relational quantities that emerge from gauge-fixing are usually non-local (with respect to the supporting manifold) because, in general relativity and Yang-Mills theory,  choices of initial data must satisfy elliptic equations, whose solutions require integrals over an initial spacelike surface.
 In practice, ellipticity means that boundary value problems require only the boundary configuration  of the field, i.e. they  do not also require the field's rate of change at the boundary. Thus the solution of these equations exists \emph{on each simultaneity surface}; and so the solution does not describe the propagation of a field, as would a solution to a hyperbolic equation. Thus non-locality may arise because the function that takes the original local degrees of freedom to a uniquely propagated subset  is, generally, non-local: the value of an element in the subset at point $x$ depends on the values of the original degrees of freedom at other points.  

But this non-locality can also often be understood as emerging from relations to a reference frame: for instance, geodesic distance relative to a boundary in a geodesically convex region, would instantiate non-locality, and would also fix the spatial diffeomorphisms of the initial surface relative to fixed boundary cf. \cite{GomesRiello2016,carrozza2021, carrozza_edge_2024}. Once we describe all fields relative to such a reference frame, they are better seen as `composite' or `dressed' by functions of all of the fields: they can be then expressed as gauge-invariant functions.     

But we can bypass the mention of reference frames and obtain a straightforward  equivalence between gauge-fixings and invariant, composite or `dressed' functions by following \cite{Rep_conv}.

We quickly summarise the method. Given a gauge-fixing function $\mathcal{C}$, there is a gauge transformation $\gamma$ that takes any given configuration $\varphi$ into one satisfying the gauge fixing, $\mathcal{C}=0$. The mapping $\varphi\mapsto\gamma(\varphi)$ is then referred to as \textit{the dressing}, and in the case of gauge-fixing functions that involve derivatives it usually involves integrals and is therefore non-local. One can then verify that the \textit{dressed} fields, $\varphi^{\gamma(\varphi)}$ are gauge-invariant quantities.

Now, one can also understand the dressing $\gamma(\varphi)$ as implicitly tied to a reference frame as follows: it is the transformation---e.g. the diffeomorphism---that takes an arbitrary representation of the state to the representation according to the reference frame.

To illustrate, suppose we have four scalar fields that are linearly independent in some region; say, four solutions of Klein-Gordon equations for different initial values, call these $\phi^{(I)}$, where $I=1,...,4$. Thus, putting the four together into a four-tuple, we have local diffeomorphism, for $U\subset M, \bar U\subset \mathbb{R}^4$, $\phi^{(I)}:U\rightarrow \bar U$. Of course, since these are physical fields, they will appropriately covary under the action of the diffeomorphisms: this is the property required of $\gamma$ above. Now, using this map, we can write: $g_{IJ}(\phi):= \big[(\phi^{(I)})^{-1}\big]^* g_{ab}$, where with $\big[ \cdot \big]^*$ we denote the pullback and with $g_{IJ}(\phi)$ the components in the frame $\{\phi^{(I)}\}$ of the abstract metric tensor $g_{ab}$. Such a quantity is commonly defined a \lq{}\textit{relational observable}\rq{} and it is clearly invariant under the active diffeomorphisms. See \cite{goeller_diffeomorphism-invariant_2022,Kabel_et_al2024,Rep_conv, BamontiRF, BamontiGomes_RF} for more about such examples and their relation to gauge-fixings and dressings).  
\ 

In addition to these facts about the dynamics of gauge theories, we can give two further points about the construction of invariant observables:\\
(i) 
Both gravitational or material degrees of freedom can be 
employed in the construction of reference frames/dressings/observables;\\
(ii) Reference frames, or a choice of relational observables, etc, will in both general relativity and Yang-Mills theory only be available \textit{locally} in field space. In the gauge-fixing language, this is a consequence of the \textit{Gribov obstruction} \cite{Gribov:1977wm, Singer:1978dk}. Conceptually, the obstruction emerges from the physical nature of the reference frame: for some field values, the configuration that made up the reference frame simply won't have sufficient structure to serve its role as a reference frame. 

Let us add a couple of remarks. The non-locality we have been referring to in the above discussion is with respect to the manifold supporting the fields. This non-locality could be seen as a further suggestion that manifold points have indeed no direct physical significance, albeit they may enter indirectly in the construction of physical (diffeomorphism-invariant) quantities. Indeed, in the relational strategy, physically meaningful spacetime localization is always with respect to field values, i.e. those fields used as reference frames, and local physics is actually to be defined with respect to such field values\cite{goeller_diffeomorphism-invariant_2022}. One could say that local physics takes place in field space, not the supporting manifold. This is one motivation to consider dynamics in such field space as the relevant arena for extracting physical effects from quantum gravity models \cite{OritiHydro}.

\

\textbf{3.}
\textbf{Physical frames are (very) different from coordinate frames:}

The conclusion of the relational strategy for the construction of diffeomorphism-invariant observables and a physical definition of spacetime localization is that it  requires the use of physical reference frames, constituted by dynamical entities (fields) in the model. Obviously, this is not only in line with the formal requirements of the gauge invariance, but also with the lesson of background independence. 

It is important to realize that this step brings truly novel features in our models of gravitational systems, and not just a formal adequacy to mathematical requirements. Physical reference frames possess properties that coordinate frames do not possess, and such properties have to be carefully accounted for in our models, otherwise we will pay the price of missing out a number of potentially crucial physical effects.

Let us illustrate these properties by stressing the difference with coordinate frames. 

Physical frames, in general, can be adopted only \lq locally\rq, i.e. for a limited range of values of the fields that we interpret as representing temporal and spatial distances, just like coordinate charts in general only make sense locally on the manifold. As with standard coordinate charts, this is markedly the case on manifolds with non-trivial topology.  But, as we mentioned above when discussing the Gribov ambiguity,  unlike coordinate charts, the limitations of applicability of physical reference frames can also be due to dynamical considerations, eg a certain set of fields can be only used as a reference frame for a subset of solutions of its equations of motion, or for a restricted region in field space.

More crucially, since physical reference frames are themselves fully dynamical, they must interact with other dynamical entities in the model. In particular, they are coupled to geometry via Einstein's equations (or those of other diffeomorphism-invariant gravitational theories), thus they back-react on the geometry in a way that depends on their own energy-momentum tensor. This back-reaction may not be negligible. 

In general, these and other physical effects affect the gravitational dynamics of spacetime and matter described in terms of physical reference frames \cite{Giesel:2020xnb}.

The above is the most immediate (interesting) complication
arising from using physical frames, in the classical domain. Being truly physical, however, dynamical reference frames are also quantum systems. 

This means that, in a truly realistic model of our gravitational systems, accounting for diffeo-invariance via the relational strategy, our clock and rods should be described as any other physical system in the model, with their own Hilbert space of states, their algebra of observables etc. It is intuitively clear that this brings a whole new set of complications, which are not merely formal, and which are being carefully addressed in recent research (see \cite{Kabel:2024lzr,delaHamette:2020dyi,hoehn_trinity_2021} and references therein). To list a few: quantum reference frames will generically be in a superposition of the states corresponding to definite values of the observable we identified as our clock or rod reading, thus such reading will be subject to quantum fluctuations; quantum reference frames will be entangled with other physical systems; quantum reference frames will be subject to contextuality conditions; typical notions like the relative state of the system being in a superposition or carrying entanglement between subsystems depend crucially on which reference frame's point of view is used.\\ 
In particular, as soon as one considers \textit{multiple} observables relating to a given frame, one cannot in general make all of them sharp in a given state (as opposed to a single observable, where this is always possible). The above is generic, but it has immediate implications for spacetime localization. For example, we may expect that incompatibility conditions will affect those observables that we define such spatiotemporal localization (generically requiring multiple observables), resulting in some form of spacetime non-commutativity. Generally speaking, the proper use of quantum reference frames in gravitational physics has been proven to inform and clarify our notions of localiseability\cite{giacomini_spacetime_2021,chataignier_relational_2024}, time evolution\cite{hoehn_matter_2023,hausmann_measurement_2024}, gravitational entropy\cite{vuyst_gravitational_2024} and the wave function of the universe\cite{hoehn_switching_2019}.

Moreover, quantum uncertainty could make it unreasonable to neglect the backreaction of physical frames on the geometry. 


Take the simple example of a scalar field $\phi$ used as a clock, in a Hamiltonian framework. Its conjugate variable, the momentum $\pi_\phi$, can be understood as a contributor to the energy density of the field.\footnote{Of course, in suitably engineered situations, one might devise a clock whose conjugate variable only indirectly couples to gravity. The general issue, however, remains.} 
Therefore, at a classical level, one must keep said momentum as small as possible, in order to allow for a suitable evolution of the value of the clock field while keeping gravitational backreaction to a minimum.
However, if the field is quantum, this is no longer an option. Due to the uncertainty principle, there is complementarity between the two variables and one cannot specify them both in \textit{any} state to an arbitrary degree of sharpness. Particularly, naively making the determination of the clock value sharper and sharper induces large uncertainties in the value of the momentum. This means that it is no longer accurate to neglect the backreaction caused by the field (because a small expectation value of the clock value is overwhelmed by fluctuations, which involve possibly quite large energy densities participating in the quantum interaction of the clock with the gravitational field, and which could induce similarly large fluctuations in geometry too).

All the above has to be taken into account in models of quantum gravity where gravitational dynamics is studied (and extracted) in relation to physical frames \cite{MarchettiOriti}.

\

\textbf{4.}
\textbf{Coordinate frames are physical only in (useful) idealizations, which have to be ultimately removed}

Coordinate frames are the result of an idealization (which is technically implemented by various approximations). They correspond to specific reference frames used by agents/observers, and can be obtained as a limiting case of physical reference frames modelling them (using the terminology explained above),
and thus physical degrees of freedom, in the idealized case in which they are modelled as not gravitating, not interacting with other physical systems and classical \cite{Dittrich,GieselThiemannScalar}. Removing this idealization requires one to further develop 
the construction and analysis of physical (diffeomorphism invariant, relational) observables in classical and quantum gravity, and 
the construction and analysis of quantum reference frames. It also requires gaining proper control over the coupled dynamics of such physical reference frames and other systems, including spacetime itself.

The success of the routine use of coordinate frames in physics (including gravitational physics) shows that the approximations leading to idealized frames are applicable and robust in most regimes of interest. The point about them being idealized remains however, and we should consider carefully and critically such approximations in the context of models in which they are not assumed from the start: we could be missing interesting new physics. In particular, in regimes in which quantum gravity is relevant, we should not expect such approximations to be valid, in general.

There are two cases, however, in which standard coordinate frames can be physical and diffeomorphism invariant, i.e. in which it is not apparent from the mathematical structure of the model that these frames are the result of a drastic idealization from a more appropriate modelling of physical reference frames (used by the agent/observer). In the first, the spacetime geometries being considered possess special isometries. They thus carry a notion of preferred observers and spacetime directions. In the second case, the gravitational system is modelled in terms of spacetime with boundaries at infinity and corresponding special asymptotic boundary conditions. While these situations are indeed special cases, and one that corresponds to the idealized case of neglecting entirely quantum features, and therefore do not alter the conceptual significance of the above conclusion, which applies to the general case, it is important to realize that another crucial idealization is in place in the second type of situation as well. 

Indeed, to assume that the gravitational system under consideration occupies an infinite region of spacetime, and that the observer studying it can sit at an asymptotically safe distance from it, is, clearly, another (useful) idealization. It corresponds to the case in which we consider a system that is actually localized in some finite spacetime region that just happens to be big enough relative to physical reference frame to be idealized as infinite.   

Removing this idealization and thus making our models of gravitational systems more realistic, means tackling a number of challenges and further developing  a research direction that has recently gained much attention: that of gravitational physics in finite regions, and the novel features that are brought into relevance by the presence of finite boundaries. 

A general result is that such boundaries should be endowed with dynamical edge modes, also in order to preserve full gauge invariance, when diffeomorphisms are taken to act on boundary degrees of freedom. 
A second result is that edge modes of such regions can be seen as providing a physical reference frame, so that defining a physical spacetime boundary in all its dynamical aspects is equivalent to selecting a physical frame.  These two results form the basis for the next two steps in our argument. 

\

\textbf{5.}
\textbf{Considering gravitational physics in finite regions brings further physical features}

  Gauge symmetries in  a bounded subsystem are quite subtle. Dynamical structures for the subsystem, such as its intrinsic Hamiltonian, symplectic structure, and variational principles in general, acquire terms that must be carefully dealt with.

The  conceptual reason for the subtle treatment of gauge symmetry in subsystems is that gauge theories manifest a type of non-locality, as we described above. Thus the global, physical phase space (or the corresponding global physical Hilbert space, in the quantum theory) is not factorizable into the physical phase spaces over the composing regions.\footnote{This type of holism, or non-locality is a well-known issue for theories with elliptic initial value problems: e.g. Yang-Mills theory and general relativity. For a reference that explores non-factorizability in the context of the holonomy formalism,  see \cite{Buividovich_2008}. For the relation between different symmetries and locality in gauge theory, see \cite{Elements_gauge}, and for the non-locality of gravitational invariants see e.g. \cite{Torre_local, Donnelly_Giddings}; for more recent use of this non-factorizability in the black hole information paradox, see \cite{Jacobson_2019}.  For a discussion of the relation between the factorizability of Hilbert spaces and the augmentation of the phase space with `edge-modes', which we will shortly discuss, see \cite{Teh_abandon, Geiller_edge2020,  carrozza2021}. \label{ftnt:edge_nonlocal}} 
 
The technical reason is that, setting aside very stringent physical boundary conditions---such as vanishing field-strength in the case of Yang-Mills theory---the variation of the action functional produces boundary terms that are not gauge-invariant. This is easy to see: any Lagrangian that employs variables that are not gauge-invariant---and this includes both general relativity and gauge theory---will, upon variation, produce boundary terms that depend on the variation of these variables. On the initial value surface, such terms are innocuous, and will merely inform us of the symplectic structure of the theory: which variables are canonically conjugate. But for timelike boundaries, we obtain non-invariant terms. How should we interpret these?

We can easily illustrate this with the  Yang-Mills action in vacuum: (capital Roman letters referring to Lie algebra indices)  
\be S(A):=\int_{M\times \RR}   F^I_{\mu\nu}F_I^{\mu\nu}.
\ee
On a bounded submanifold, say, $R\times \RR$, where $R\subset M$ is a spatial submanifold of $M$ and $S=\partial R$, a variation of the action yields, after integration by parts: 
\be\label{eq:deltaS} \delta S(A)= -\int_{R\times \RR} \delta A_I^\nu(\D^\mu F^I_{\mu\nu})+\oint_{S\times \RR} s^\mu  F^I_{\mu\nu} \delta A^\nu_I,
\ee
where $s^\mu$ is the normal to the hypersurface $S\times \RR$ in ${M\times \RR}$. 
 Now, for the first term of \eqref{eq:deltaS} to vanish for arbitrary variations of the gauge potential it suffices that the gauge potential satisfies the Yang-Mills equations. But the second term vanishes only if either (the normal component of) the  field tensor  vanishes along the boundary or $\delta A^I_\mu$ vanishes at the boundary. The first condition is severely limiting; the second is not a gauge-invariant condition.\footnote{ It is important here that these are time-like boundaries; for the spacelike initial and final surfaces, one can implement whatever initial conditions one likes. For this reason, most other work deals with non-covariance in the Hamiltonian or symplectic formalism. And the boundary term gives rise to the symplectic potential: $\theta=\int E^i\delta A_i$, which defines the symplectic structure of the theory, $\Omega:=\delta\theta$.  This is done with much greater care in \cite[Secs. 4-7]{carrozza2021}.  
 } 

The standard approach to non-asymptotic spatial boundaries treats the lack of  invariance of the subsystem 
by paring 
down gauge symmetries at the boundary.
 


However,
the recent flurry of papers on subsystems in gauge theory, 
calls the pared-down treatment of internal boundaries of subsystems into question  (cf. e.g. \cite{Teh_abandon, Geiller:2017xad, DonnellyFreidel, GomesRiello_new, Riello_symp, GomesStudies, Geiller_edge2020, carrozza2021} and references therein). New mechanisms, for instance, `edge-modes', have been devised to maintain the gauge invariance of the internal boundary under all gauge transformations (e.g. even ones with support on the boundary).
By edge modes, generally speaking, we mean sets of degrees of freedom that are localised on boundaries only, and particularly any physical reference frames that naturally can be constructed from them.

\

\textbf{6.}
\textbf{Edge modes provide/allow for construction of physical frames.}\\
 Using representational schemes consisting of physical reference frames or gauge-fixings, one can find gauge-invariant regional dynamical structures labeled by the (equivalence classes of the) fluxes at the boundary; see \cite[Sec. 4]{Riello_symp} and \cite[Sec. 3]{GomesRiello_new}. In a similar fashion, in \cite{carrozza2021,carrozza_edge_2024} these dynamical structures are obtained by stipulating a representational scheme solely at the boundary; they call such conventions  \emph{boundary reference-frames}.

Although these approaches vary to a certain extent, here we can try to describe their common core. The main idea is to find a physical representation of the symmetry-variant variables, e.g. which we called $h(\varphi):=\varphi^{\gamma(\varphi)}$, and correspond to the projection of the field along the corresponding gauge-fixing surface. Variation will then act not only on the bare variable $\varphi$, but also on the dressing, $\gamma(\varphi)$. This variation cancels the troublesome non-gauge-invariant boundary terms. 

In a bit more detail, here is the rough idea  why this works (see especially \cite{GomesRiello2016, GomesStudies}): the problem with the extra term in the variation of the action in \eqref{eq:deltaS} 
is essentially a lack of gauge-covariance under the variation $\delta$. 

To correct for this lack of covariance, we can use analogies between finite and infinite-dimensional geometry. Using these analogies, we can relate $\delta$ and spacetime derivative operators, $\d$. And similarly, for theories with symmetries, there is  an analogy between the infinite-dimensional space of models $\F$ and finite-dimensional principal bundles, $P$. In the finite-dimensional case, to get around a lack of gauge-covariance under spacetime-dependent gauge transformations, we introduce the connection $\omega$ on $P$ through minimal coupling, i.e.  so that, schematically, quantities acted on by the \emph{horizontal} derivatives $\d\rightarrow\d_h:=\d-\omega$ (where $h$ stands for `horizontal')  remain covariant under spacetime-dependent gauge transformations.  So  too, in the infinite-dimensional case, we can justify the introduction of a connection-form $\varpi$, such that $\delta\rightarrow \delta_h:=\delta-\varpi$ becomes a fully covariant operator. These `connection-forms' are the infinitesimal versions of the dressing function $\gamma(\varphi)$.
That is, the $\varpi$ generalise the dressing of the fields that one would obtain via gauge-fixing to an infinitesimal setting; because they are (group-valued) functionals of the fields, they are relational constructions. Each choice of $\varpi$ gives a decomposition of field variations in terms of `pure gauge' and `physical': but again, what is `physical' depends on a choice.\footnote{In the language of \cite{Rovelli_partial}, they become \textit{partial observables}. The subscript $h$ is appropriate since $\varpi$ is associated with a horizontal projection operator on $\F$.  The difference between a horizontal projection operator mentioned here and a projection onto a gauge-fixing surface is that the former only needs to be `distributional': it acts as a projection only within each $T_\varphi\F$. When such a distribution is integrable, meaning that the associated infinite-dimensional connection form $\varpi$ has no associated curvature, then the horizontal spaces foliate the space of models, and each leaf will correspond to the range of a particular choice of $\sigma$, and each initial value  selects an entire leaf: this will be equivalent to a gauge-fixing.  The infinite-dimensional connection-forms  generically have curvature; but the curvature vanishes in the Abelian theory, i.e. electromagnetism (cf. \cite{GomesHopfRiello} for a comprehensive account).  \label{ftnt:hh}}

 In a similar spirit,  \cite{carrozza2021,carrozza_edge_2024} 
 demand a decoupling of the variational action principles of complementary spacetime regions separated by a timelike boundary (as in \eqref{eq:deltaS}). This is done in two steps: (i) restricting the state space by physical boundary conditions, such as fixing the electromagnetic field at a boundary, i.e. fixing a superselection, or, in their words, post-selected sector; and (ii) adding boundary terms to the action functional so that, within the physically restricted state space, the regional variational principles---such as in \eqref{eq:deltaS}---fully decouple. 
As we have described in this Section, point (i) has a subtlety:  in  gauge theory,  it is a non-trivial matter to fix physical, or gauge-invariant, boundary conditions. To accomplish  that, \cite{carrozza2021,carrozza_edge_2024} help themselves to  reference frames at the boundary. 
Relative to these reference frames we can fix gauge-invariant boundary conditions for the gauge potential or the metric. 
 

 The resolution of the failure of gauge covariance at the boundary is pursued differently in \cite{DonnellyFreidel} and follow-up papers (see e.g. \cite{Geiller_edge2020} for a more complete list).   These papers add new degrees of freedom at the boundary with appropriate gauge-variance properties so as to cancel out the unwanted terms. In some circumstances the two approaches are related through a suitable interpretation of the new degrees of freedom: (see e.g. \cite[Section 5]{Riello_symp}, \cite[Section 5]{ReggeTeitelboim1974} \cite[Section 4]{carrozza2021}, and \cite{Teh_abandon}). But many find the introduction of new degrees of freedom problematic; (see \cite[Sec. 7]{GomesRiello_new}, \cite[5.6]{Riello_symp} and \cite[Sec. 3.2]{GomesStudies} for critiques).

\

\textbf{7.}
\textbf{We are left with only physical reference frames, but a more realistic (and closer to operational) local physics}

The end result of removing both kinds of idealizations, i.e. that of frames being expressed by coordinate choices and that of infinite spacetime regions, leaves us with models of gravitational systems defined on (collections of) finite regions of spacetime with geometric (thus gravitational) observables expressed as relations between field values (including the metric field).

Of course, it is difficult to explicitly realize such constructions. Classical deparametrization of GR in terms of material frames has been achieved globally only in terms of not-entirely-physical material systems. A patchwise construction of generic gravitational dynamics with realistic matter frames is yet to be developed. Quantum reference frames, as studied so far, neglect entirely the gravitational coupling and most dynamical aspects. Constructions within full quantum gravity, thus both fully dynamical and fully quantum, have been only achieved in spatial approximations \cite{MarchettiOriti}. Quantum gravitational toy models, such as those employed in quantum cosmology in which relational dynamics is extensively studied, are indeed just toy models.\\
Despite all the above limitations, thus the lack of explicit, complete and fully general realizations of a description of gravitational physics in a diffeo-invariant language entirely in terms of physical reference frames, the above arguments and known partial results convince us that this is what we should strive for, to achieve a more realistic modelling of gravitational systems and spacetime physics. We now expand and draw some further implication from this conclusion. 

\

\textbf{8.}
\textbf{Gravitational (spacetime) physics is intrinsically perspectival}

As we have argued, a proper modelling of a (quantum) gravitational system must invoke descriptions involving relations between subsystems, and such relations are expressed relative to physical reference frames used by some general kind of observers. 
On the one hand, such a style of modelling is incentivised by the ambition to be realistic in our understanding of physics, but as we have argued, it is also made an implicit necessity by the formalities of gauge theories, in particular gravity.
In general, now, these reference frames may be of totally different structure, validity and quality. Therefore, their corresponding descriptions of a system 
are a priori very different as well. \\
To illustrate this idea, in its simplest form, we can point to the various different types of ways one can implement a physical clock. The perhaps most precise type uses atomic orbital transitions, but it is also possible to use something more elementary like a pendulum. In a gravitational context we can expect the two to behave qualitatively quite differently, as one of them makes explicit use of its gravitational interaction. However, in principle the atomic clock also couples gravitationally, and is bound to have a limited range of validity as a reliable reference frame. Therefore, the coupling of the reference frame and the remainder of the system is sometimes negligible, sometimes non-negligible, and sometimes even essential for its interpretation as a frame. \\
Now, according to  a realistic account of agents and their physical reference frames, any set of relational observables is expressed, by definition, in the perspective of some agent participating in the dynamics of spacetime. Given this issue, one has already on the purely formal level an attachment of modelling activity and its results (predictions etc.) to the concrete perspective of some agent. \\
Furthermore, existing models for these situations, such as different choices of physical clocks in quantum cosmology  \cite{Gielen:2021igw,Giesel:2023pfl}, show that even at a practical level, this attachment can be a nontrivial issue. 
In a quantum context, there are even further complications beyond the classical regime. While classically, the relational strategy  provides a resolution to the  lack of diffeomorphism-invariant observables, their quantum versions are less useful in identifying conventional objects (like spacetime events) familiar from local nonrelativistic physics \cite{Kabel:2024lzr}. Such complications then further emphasize that the choice of reference frame directly correlates with an appropriate set of accessible observables.\\
We, therefore, have to assert, when no further qualification is present, that concrete physical models associated with different physical frames are, in principle. unrelated. Thus, since the description is contingent on a physical frame whose qualities influence the modelling process, we can call it  \textit{perspectival}.

\

\textbf{9.}
\textbf{We may have no general covariance, not to mention invariance, in the more realistic, physical context}

At this point, it should be clear that for any model of gravitational physics (as well as gauge physics), we have at our disposal, in principle, several possible physical frames.  But the standard notion of general covariance (the standard independence of non-physical coordinate charts or diffeomorphism invariance) of general relativity  is made irrelevant (because already accounted for) by the use of physical frames in the relational framework.
Depending on their physical (dynamical) properties, including quantum features,  one frame may be obviously more practically useful than another, but in a realist reading of the theory, this practical aspect would not point to a more fundamental ontological status for one frame over another.  

 The question arises, then, of whether we have any new notion of \textit{physical} general covariance among the physical reference frames. That is, are there transformation maps between the different relational descriptions defined in terms of different physical frames, and what, if anything, is conserved or invariant under such mappings? 
 
 One could argue that some form of physical covariance is expected, and we have both general mathematical arguments as well as explicit examples of such physical covariance, which are robust at the classical level. These examples usually involve using the redundant parametrized formulation written in terms of diffeo-covariant quantities only as a bridge between two different deparametrized formulations in terms of diffeo-invariant relational descriptions in terms of two different physical (material) frames. Such examples also usually involve going partially on-shell with respect to some equation of motion, limiting their generality and their applicability beyond the strictly classical context. 
 
 On the other hand, even if full physical covariance could be achieved, one should not expect, generically, invariance of all physical properties or dynamical features under the identified maps between different relational descriptions corresponding to different physical frames. Such maps would not correspond to gauge symmetries and so we would lack any mathematical reason to enforce invariance. Again, we have examples pointing to these types of exceptions. 
 
 As a relevant further complication, we have the issue that our understanding of general reference frame transformations at the quantum level is still a work in progress. Available results concern both simplified settings \cite{Krumm:2020fws,Vanrietvelde:2018dit} as well as more generic constructions\cite{hamette_perspective-neutral_2021}. The possibility of switching between quantum reference frames in a full dynamical context passing though a (redundant) perspective-neutral formulation, like in the classical case, is a promising avenue \cite{Vanrietvelde:2018pgb,hamette_perspective-neutral_2021}, and has been argued to be equivalent to, or even subsume other, more conventional approaches like the Page-Wootters formalism\cite{hoehn_trinity_2021,hoehn_equivalence_2021}. 
 
 In the cases where such an understanding is under control, such as that of finite dimensional quantum reference frames, we witness several salient, interesting features. Important properties of quantum systems like superpositions and entanglement of subsystems become, to a notable degree, perspectival. The same state of system and agents, as viewed from one particular agent's perspective, may be sharp or entangled in a certain observable, or it may be perfectly smeared out or disentangled in another agent's perspective (cf. \cite{Kabel_et_al2024} for examples). While this may simply hint at more complicated rules of quantum covariance\cite{cepollaro_quantum_2024}, it is striking as it implies the sharpness of the frame---as linked, by the argument we made earlier, to the strength of backreaction---is further contextualised. Sharpness may be relative to the view of the system from an agent's perspective. Flipped on its head, taking the perspective of \textit{any} one agent makes their reference frame (including any boundaries) sharp; but this then leads to a backreaction with possibly infinite strength due to fluctuations in conjugate variables. Therefore, quantum effects also put limitations on how strictly one can switch from one observer/agent's perspective to another. 
\

At the end of these steps, we are thus left with more questions than answers, about what we obtain after removing all the mentioned idealizations from our models of gravitational systems and spacetime physics. The main conclusion is that we should be left with a description fully formulated in terms of relational quantities associated to our choice(s) of physical reference frames, and that we should strive to achieve such description in greater generality, beyond the simplified or otherwise unsatisfactory examples we have currently at our disposal. We are also left with the fundamental questions of how to link the different descriptions associated to different choices of such physical reference frames, and of what, if anything, is invariant across them.

\

\section*{Discussion}

In this contribution, we recalled ideas and results from three research directions, gravitational physics in finite regions, relational dynamics in classical and quantum gravity, and quantum reference frames, to advocate for a description of spacetime physics formulated fully in terms of physical reference frames: physical systems used as clock and rods to localize other physical systems in space and time, and modelled without neglecting their physical features. We argue that this is the result of removing two types of idealizations from our physical models: that corresponding to idealized reference frames coded in coordinate systems or gauge fixings, and that corresponding to closed or asymptotic boundary conditions. We emphasized that it is a partially open issue to understand what would be, in such description, a fully physical counterpart of the standard general covariance of gravitational physics with respect to changes of coordinate frames, i.e. a covariance with respect to transformations across physical reference frames. Moreover, we have emphasized that we have no reason to expect, even assuming that such physical covariance can be established, that physical properties would remain invariant under the same set of transformations. The dependence of the \lq facts about any given gravitational system\rq ~on the chosen physical reference frame would instead be a feature to accept as fundamental and to understand in its full implications. The conclusion is that removing standard idealizations from gravitational physics makes a more fundamental (quantum) understanding of it necessarily perspectival, in the sense of being always relative to a particular choice of {\it physical} reference frame. This perspective is already present in a number of works in the literature and we emphasize its foundational importance. 

\

Before we add some further comments about our arguments and conclusion, and their possible conceptual implications, we want to emphasize their generality.

For concreteness, we have mostly focused on the gravitational case. Beyond its foundational significance and universality (any physical system is localized in spacetime, except spacetime itself, and interacts gravitationally), this case has the advantage of tying closely to our intuition. We have an intuitive understanding of what clocks and rods are, and what could model (and generalise) their functions. But most of our arguments and all the conclusions we have reached apply to any gauge theory. The recent work (and most results) about field theories in finite regions, edge modes and boundary charges has been developed for both gravitational physics and any other gauge interactions, in particular non-abelian gauge theories. The role of edge modes in preserving gauge covariance in finite regions as well as their understanding as defining physical reference frames at the boundary of such regions remain valid. Moreover, any gauge symmetry can be construed geometrically (cf. \cite{Gomes_internal}), and the explicit inclusion of physical reference frames in our models are also necessary in gauge theories for gauge-invariant descriptions based on a relational strategy. The main difference is that such reference frames typically concern internal vector spaces, as opposed to spacetime, but this distinction does not affect the relevance of our arguments. These arguments appear, therefore, even more general than what we discussed.

\

Let us now discuss some more conceptual aspects of our conclusion about the importance of investigating further the issue of physical general covariance, and the dependence on any given choice of physical reference frame. 

These conceptual aspects, on the one hand, motivate further work linking this issue in the foundations of gravitational physics with other work in the foundations of quantum mechanics; on the other hand, they point to important open issues of a more general philosophical nature, that deserve as well further analysis and a proper treatment (which goes, however, beyond the scope of this contribution).


First of all, even if based on a large number of existing solid results, and regardless of how much we may be personally convinced, our conclusion is, to a large extent, mostly a pointer and an encouragement to further research. Each of the three research directions our argument builds on is a work in progress, with much left to be understood or established more rigorously. Some important open issues are: the quantum counterpart of the recent work on gravity in finite regions and edge modes; the definition and control over (relational) spacetime observables within quantum gravity approaches; the extraction of an effective gravitational dynamics within the same approaches; the extension of treatments of quantum reference frames to the dynamical gravitational context; the analysis of symmetries in field space that could represent (or contribute to) a physical counterpart of general covariance at both classical and quantum level, to name but a few.  

Next, we argued that physical reference frames (and reference frames more generally) can be seen as a partial embodiment, in our physical models, of observers and of the very epistemic agents constructing/using such models. In the same direction, the removal of both kind of idealizations of idealized frames and of infinite or closed spacetime regions, which leads, we argued, to a description of gravitational systems that is necessarily relative to such physical frames, have to do with taking more seriously the role of observers and the open nature of physical systems. It is the role of observers and the system/observer split that have to become central in our understanding of physics; this is our conclusion, in more general terms, abstracting from our specific gravitational context.

From this more general perspective, there are other research directions that can be seen as supporting our general conclusion but which we did not mention. In particular, we have in mind the issues raised by Wigner's friend scenario and its many modern generalizations, for example, the Frauchiger-Renner thought experiment \cite{Frauchiger2018}. In our reading, these scenarios suggest, in different ways, that indeed quantum physics may be intrinsically perspectival and that (quantum) facts are ultimately relative, in subtle and interesting ways \cite{Brukner2020, Brukner2018}. In turn, this aligns with a number of interpretations of the quantum formalism (pragmatist, relational, subjective, in different forms \cite{Barzegar:2022evv, Cabello_2017, Healey2012, sep-quantum-bayesian, Dieks_2022}) that are based on a relational, when not epistemic, approaches to quantum states and observables.

Following this chain of thoughts, thus taking seriously the link between physical reference frames in a gravitational context and observers (or epistemic agents), we can suggest further implications of our arguments, of an epistemological kind. 

The epistemological counterpart of (physical) general covariance can be taken to be complete intersubjectivity. By this, we mean the possibility of sharing fully, by appropriate translation across the perspectives associated to individual observers or epistemic agents, the content of each perspective, not just approximately but exactly, not only part of the physical content of each perspective but all of it. In other words, different observers are able to recognize what, exactly, physical facts and properties are, with respect to each of them.

The epistemological counterpart of complete invariance under change of (physical) reference can be taken to be a stronger notion of objectivity. By this, we mean a complete agreement, across observers (or epistemic agents), about physical facts and the properties of physical systems. In other words, different observers are not only able to recognize what, exactly, physical facts and properties are with respect to each of them, but they actually agree on such facts and properties. 

In this vein, the classical and quantum gravitational challenges to physical general covariance, which, we argue, is yet to be fully established must be investigated in more detail because of its central importance, are then challenges to intersubjectivity. And the fundamental limitations we suggest for the existence of complete and exact invariance under changes of physical reference frames are then fundamental limitations from gravitational and quantum physics to objectivity.

Thus, with the above identifications in mind, our physical reasoning is inspired by and lends support to the many philosophical viewpoints (in the philosophy of physics and philosophy of science more generally) that have advocated for some perspectivalism in our epistemology, for a weakening of the notion of attainable objectivity, and in general for a greater role for agents and observers (some examples are \cite{Giere2006-GIESP,Massimi2017-MASPNC,Brown2009-BROMAP-10, VanFraassen2008-VANSRP-2}). 

It is clear that, at the present stage, this conclusion is not much more than a(nother) pointer to further research, this time in the philosophical domain; and that proper philosophical analysis if called for, in order to put these suggestions on solid ground. We hope, however, that the foundational interest of such an analysis for physics and philosophy is also clear.

\section*{Ackowledgements}
We thank one anonymous referee for several useful comments on the first version of this contribution.
DO acknowledges  support through the Grant PR28/23 ATR2023-145735 (funded by
MCIN /AEI /10.13039/501100011033). SL thanks E. Telali for pointing out that single observables can still be made sharp in QRFs, B. Sahdo for the two examples of contrasting clocks and thanks the Perimeter Institute for hospitality. Research at the Perimeter Institute is supported in part by the Government of Canada through NSERC and by the Province of Ontario through MEDT. SL and DO acknowledge funding from the Munich Center for Quantum Science and Technology. HG acknowledges the British Academy for support.


\bibliography{references3,essaz}
\bibliographystyle{apacite} 
\end{document}